\documentclass[10pt,conference]{IEEEtran}
\usepackage{enumerate}
\usepackage{amsthm}
\usepackage{ragged2e} %To justify text
\usepackage{amsmath}
\usepackage{amscd}
\usepackage{dsfont}
\usepackage{amssymb}
\usepackage{latexsym}
\usepackage{array}
\usepackage{url}
\usepackage{hyperref}
\usepackage{tabularx}
\usepackage{tikz}
\usepackage{hyperref}
\usepackage{graphicx,subfigure}
\usepackage{bbm}
\usepackage[autostyle]{csquotes}
\usepackage{hhline}
\usepackage[numbers, square, comma, compress]{natbib}
\usepackage{multirow}
\usepackage{nameref}
\newcolumntype{C}[1]{>{\centering}m{#1}}

\newtheorem{lem}{Lemma}
\newtheorem{corol}{Corollary}%[section]
\newtheorem{ther}{Theorem}

\newtheorem{prop}{Proposition}

\theoremstyle{definition}
\newtheorem{rem}{Remark}
\newtheoremstyle{dotless}{}{}{\itshape}{}{\bfseries}{}{ }{}
\theoremstyle{dotless}

\newcommand{\bm}[1]{\mbox{\boldmath{$#1$}}}

\newcommand {\aplt} {\ {\raise-.5ex\hbox{$\buildrel<\over{\mbox{\scriptsize $\sim$}}$}}\ }

\renewcommand{\IEEEQED}{\IEEEQEDopen}

\DeclareMathOperator*{\Mod}{\,mod}

\makeatletter \g@addto@macro\@floatboxreset\centering \makeatother

\begin{document}

%\title{Secret Key Generation Using Lattices}
%\title{Secret Key Generation from Gaussian sources using lattices in the presence of an eavesdropper}
%\title{Secret key generation \\ from Gaussian sources using lattices}
\title{A reconciliation approach to key generation based on Module-LWE}

\author{\IEEEauthorblockN{Charbel Saliba and Laura Luzzi}
\IEEEauthorblockA{ETIS, UMR 8051, \\ CY Universit\'e, ENSEA, CNRS,\\ Cergy, France
 \\
Email: {\{charbel.saliba, laura.luzzi\}@ensea.fr}} \and
\IEEEauthorblockN{Cong Ling}
\IEEEauthorblockA{Department of Electrical \\ and Electronic Engineering \\
Imperial College London, U.K.\\
%London SW7 2AZ, United Kingdom\\
Email: cling@ieee.org}
}

\maketitle

%\author{
%%\thanks{%
%%This work was supported in part by FP7 PHYLAWS, by a Royal Society-CNRS
%%international joint project
%\thanks{%
%C. Ling is with the Department of Electrical and
%Electronic Engineering, Imperial College London, London SW7 2AZ,
%United Kingdom (e-mail: cling@ieee.org).
%\par
%L. Luzzi is with Laboratoire ETIS (ENSEA - Universit\'e de Cergy-Pontoise - CNRS), 6 Avenue du
%Ponceau, 95014 Cergy-Pontoise, France (e-mail: laura.luzzi@ensea.fr).
%\par
%M. Bloch is with the School of Electrical and Computer Engineering, Georgia Institute of Technology, Atlanta, GA 30332, USA
%(e-mail: matthieu.bloch@ece.gatech.edu).
%}}
%
%\maketitle

\begin{abstract}
We consider a key encapsulation mechanism (KEM) based on Module-LWE
where reconciliation is performed on the $8$-dimensional lattice $E_8$, which admits a fast CVP algorithm. Our scheme generates $256$ bits of key and requires $3$ or $4$ bits of reconciliation per dimension. We show that it can outperform Kyber in terms of the modulus $q$ with comparable error probability.
We prove that our protocol is IND-CPA secure and improves the security level of Kyber by $7.3\%$.
\end{abstract}

%\begin{keywords}
%flatness factor, lattice coding, secret key generation, strong secrecy.
%\end{keywords}

\section{Introduction}
%We define the \textit{modulus to noise ratio} as the quotient between the modulus $q$ and the parameter $r$ of the error distribution $\chi$. A smaller modulus to noise ratio provides stronger concrete security against known attacks. Moreover, since all the exchanged messages are modulo $q$, the size of $q$ affects the overhead of the protocol.
Over the past few years, there have been many attractive developments in lattice-based cryptographic protocols, 
whose security is based on {worst-case} hardness assumptions, and which are conjectured to be secure against {quantum} attacks.
Thus, lattice-based primitives are a promising candidate to replace constructions based on number theoretic assumptions like RSA \cite{RSA} or Diffie-Hellman \cite{Diffie_Hellman} that are currently in use.

One of the most versatile primitives for the design of provably secure cryptographic protocols is the \emph{learning with errors} (LWE) problem introduced by Regev \cite{LWE}.
For instance it can serve for IND-CPA (Indistinguishability under chosen-plaintext attack) \cite{LWE} and IND-CCA (Indistinguishability under chosen-ciphertext attack) public key encryption \cite{peikert2011lossy}.
\iffalse
The decision LWE problem is to efficiently distinguish a polynomial number of pairs $\left(a_i, b_i \approx \langle a_i, s\rangle \right) \in \mathbb{Z}_q^n \times \mathbb{Z}_q$ from uniform ones. Here $q$ is an integer function of $n$, $s \in \mathbb{Z}_q^n$ is a uniformly random secret common to all pairs, and the vectors $a_i$ are uniformly random in $\mathbb{Z}_q^n$ and independent. After that an error term is added to the inner product $ \langle a_i, s\rangle $ which is generated using a variant of the Gaussian distribution over $\mathbb{Z}_q$, to form $b_i$.
\fi
A structured variant of LWE, the decision \emph{ring learning with errors} ($R$-LWE) was proposed in \cite{RLWE} by Lyubashevsky \emph{et al.} to allow more compact representations, which in turn inspired the authors of \cite{module-lwe} to introduce the \emph{module learning with errors} ($M$-LWE) variant. This new problem seems to offer better security guarantees and more flexibility in designing cryptographic schemes.

%It consists in distinguishing polynomially many samples $\left(a_i, a_i \cdot s +e_i \right)\in R_q\times R_q^{\vee}$ from uniform ones, where $R$ is taken to be the ring of integers of some cyclotomic number field, $R^{\vee}$ its dual, and $q=$ poly$(n)$ a prime integer. These results were extended to more general number fields and modulus $q$ in \cite{pseudorandomness}. 

%Cryptographic applications of $R$-LWE include fast encryption \cite{RLWE} and fast homomorphic encryption \cite{brakerski2014leveled}. Solving $R$-LWE is at least as hard as solving approximate SIVP on ideal lattices. %\footnote{It is still an open problem whether SIVP on ideal lattices is as hard as SIVP for general lattices, but so far there are no known attacks that exploit the ring structure.}

In \cite{PEIKERT}, Peikert introduced an efficient lattice-based \emph{key encapsulation mechanism} (KEM) based on $R$-LWE that allows two parties to share an ephemeral key that is useful for secret communications. Peikert's scheme features a low bandwidth \emph{reconciliation} technique that aims to reach exact agreement on the shared key. A practical implementation of Peikert's protocol called \textsc{NewHope} was proposed in \cite{NEWHOPE} as a candidate to the NIST challenge on post-quantum cryptography. 
In \cite{PEIKERT} and \cite{NEWHOPE}, although key generation is performed using $1024$-dimensional lattices, the reconciliation step uses 1-dimensional and 4-dimensional lattices respectively\footnote{In fact, the latest implementation of the NewHope algorithm does not use reconciliation \cite{newhopeWithoutRec}.}. 
Subsequently, Kyber appears as an alternative approach to the NewHope KEM \cite{kyber, kyber3}. Based on the hardness of Module-LWE, the authors of \cite{kyber3} construct a IND-CCA secure KEM from a CPA-secure public-key encryption scheme which achieve $165$ bits of post-quantum security, together with an error probability less than $2^{-164}$ using modulus $q=3329$. 
 
In this paper, we consider a key-generation protocol based on Module-LWE using the reconciliation technique. We consider the cyclotomic ring of degree $n=256$ as in Kyber. Compared to NewHope, we need to use a higher-dimensional lattice for reconciliation in order to generate one bit of key per dimension. We choose the $8$-dimensional Gosset lattice $E_8$ due to its optimal density and low-complexity quantization. \\ 
We show that our scheme can guarantee a smaller error probability than Kyber's, i.e. $P_e<2^{-174}$, with a smaller modulus $q=2^{11}$, using $4$ bits of reconciliation per dimension. For this choice of $q$, our scheme achieves $176$ bits of post-quantum security. 
%Alternatively, with larger modulus $q=2^{13}$, our scheme can provide very small error probability $P_e<2^{-390}$, using $2$ bits of reconciliation per dimension. 

%Moreover, our scheme can achieve $176$ bits of post-quantum security together with an error probability $P_e \leq 2^{-148}$ and a public rate of $2$ bits per dimension.  
%Although current recommendations are to keep the error probability smaller than $2^{-128}$, this may not be sufficient when transforming an IND-CPA secure encryption scheme into an IND-CCA secure one using the Fujisaki-Okamoto transform \cite{FujOka}. 
A smaller error probability is desirable to prevent leakage of information from decryption failure attacks \cite{INDCCA}, while a smaller modulus-to-noise ratio provides better efficiency and stronger concrete security against known attacks.
%Future work will focus on more practical codes admitting lower-complexity decoding algorithms.

%Our nested lattice scheme requires that we take the modulus $q$ to be a multiple of $4$ in order to obtain an integer lattice. 
We note that by choosing $q$ to be a prime number as in Kyber, one can use the Number Theoretic Transform to speed up polynomial multiplication \cite{PG,PEIKERT, NEWHOPE}. However, prime $q$ is not required for security \cite{pseudorandomness}, and power-of-two moduli have been used in the literature in \cite{SABER,lac}. These works use other methods for efficient polynomial multiplication, e.g. Karatsuba / Toom-Cook algorithms and index-based multiplication.
An advantage of choosing an even $q$ is that a dither is not required to obtain a uniform key, unlike \cite{PEIKERT, NEWHOPE}. We note that the hardness of Module-LWE has been established for general modulus $q$ \cite{algebraically}.
\subsubsection*{Organization}
This paper is organized as follows. In Section \ref{preliminaries} we provide basic definitions about cyclotomic fields, lattices, etc. In Section \ref{alg}, we introduce our key generation algorithm. In Section \ref{proba} and \ref{security}, we provide a proof that the error probability is small, and that our scheme is IND-CPA secure respectively. Finally, in Section \ref{cbd} we provide estimates for concrete security against known attacks.  
%Finally, some conclusions and perspectives are given in Section \ref{conclusion}.

\section{Preliminaries} \label{preliminaries}
In this section, we introduce the mathematical tools we use to describe and analyze our proposed scheme. 

%We write $f(N)$ $=$ $\omega\left(g(N)\right)$ if $\lim_{N\rightarrow \infty} \left( f(N)/g(N)\right)=\infty$, and $f(N)=\Theta\left(g(N)\right)$ if $f(N)=O(g(N))$ and $g(N)=O(f(N))$. Finally a variant of $O$ notation that \enquote{ignores} logarithmic factors: $f(N)$ $=$ $\tilde{O}\left(g(N)\right)$, equivalent to $f(N)$ $=$ $O\left(g(N)\cdot \log^k (g(N))\right)$ for some integer $k$.
\subsection{Lattices and Algebraic number theory}
\subsubsection*{Lattice definitions}
For our purposes, a \emph{lattice} $\Lambda$ is a real full-rank discrete additive subgroup of $\mathbb{R}^n$. Any lattice is generated as the set of all integer linear combinations of $n$ linearly independent basis vectors $\mathbf{B}=\{\mathbf{b}_1, \dots , \mathbf{b}_n \}$ in $\mathbb{R}^n$.
The \emph{Voronoi region} of $\Lambda$, denoted $\mathcal{V}(\Lambda)$ is the set of all points in $\mathbb{R}^n$ which are closest to the origin than to any other lattice point. 
A \emph{Voronoi-relevant vector} is an element $\lambda \in \Lambda$ such that $ \langle \mathbf{x}, \lambda \rangle <\| \mathbf{x} \|^2 $ for all $ \mathbf{x} \in \Lambda \setminus \{0, \lambda\}$.
%A non-zero lattice vector $\lambda \in \Lambda$ is a \emph{Voronoi-relevant vector} if there is some $x \in \mathbb{R}^n$ such that $||x||=||x-\lambda|| < ||x-w||$ holds for all $w \in \Lambda \setminus \{0, \lambda\}$.
%For a fundamental cell $\mathcal{P}_0$, any point $\mathbf{x} \in \mathbb{R}^N$ can be uniquely expressed as
%$ \mathbf{x}= \lambda + \mathbf{x}_e$ where $\lambda \in \Lambda$ and $\mathbf{x}_e \in \mathcal{P}_0$, denoted by $\lambda = Q_{\mathcal{P}(\Lambda)}(\mathbf{x})$ and $\mathbf{x}_e = \mathbf{x} \Mod \mathcal{P}_0 = \mathbf{x} - Q_{\mathcal{P}(\Lambda)}(\mathbf{x})$. 
%Note that when $\mathcal{P}_0= \mathcal{V}(\Lambda)$, the operation $Q_{\mathcal{V}(\Lambda)}(\mathbf{x})$ becomes the closest point in $\Lambda$ to $\mathbf{x}$. 
For any $\mathbf{x} \in \mathbb{R}^n$, we denote by $Q_\Lambda(\mathbf{x})$ the closest vector in $\Lambda$ to $\mathbf{x}$.
We also define the modulo $\Lambda$ operation as 
$\mathbf{x} \Mod \Lambda = \left( \mathbf{x}-Q_\Lambda(\mathbf{x}) \right) \in \mathcal{V}(\Lambda)$. We will use implicitly in our proofs the fact that $\forall \mathbf{x},\mathbf{y} \in \mathbb{R}^n$, and $\forall \lambda \in \Lambda$, $(\mathbf{x} \Mod \Lambda +\mathbf{y})\Mod \Lambda =(\mathbf{x}+\mathbf{y}) \Mod \Lambda$ as well as $(\mathbf{x} + \lambda )\Mod \Lambda =\mathbf{x} \Mod \Lambda$.
%and if the intersection~$(\mathcal{R}(\Lambda)+\lambda) \cap (\mathcal{R}(\Lambda)+\lambda')$ has measure~$0$ for any~$\lambda \neq \lambda'$ in~$\Lambda$.
% Examples of fundamental regions
%~$\mathcal{R}(\Lambda)$
%include the fundamental parallelepiped $\mathcal{P}(\Lambda)$ and the Voronoi region $\mathcal{V}(\Lambda)$.
% All the fundamental regions have equal volume Vol$(\Lambda)$.\\
%%%%%%%%%%%%%%%%%%%%%%%%%%%%%%%%%%% EQUATIONS
% \begin{propt} 
% $\forall \mathbf{x},\mathbf{y} \in \mathbb{R}^n, \; \forall \lambda \in \Lambda$ %$ , \text{and } \forall \alpha \in \mathbb{R}$ 
%we have:
% \begin{flalign}
  % \label{eqn2}
  % &Q_\Lambda(\alpha \cdot \mathbf{x})=\alpha \cdot Q_{\frac{\Lambda}{\alpha}}(\mathbf{x}).\\
  % \label{eqn3}
 % &(\alpha \cdot \mathbf{x}) \Mod \Lambda = \alpha \cdot \left(\mathbf{x} \Mod \frac{\Lambda}{\alpha}\right).\\
%  \label{eqn4}
%  &(\mathbf{x} \Mod \Lambda +\mathbf{y})\Mod \Lambda =(\mathbf{x}+\mathbf{y}) \Mod \Lambda.\\
%  \label{eqn5}
%  &(\mathbf{x} + \lambda )\Mod \Lambda =\mathbf{x} \Mod \Lambda.
%\end{flalign}
%\end{propt}
%%%%%%%%%%%%%%%%%%%%%%%%%%%%%%%%%%% EQUATIONS
\begin{lem} \label{perm lemma}  \normalfont
Let $\Lambda' \subset \Lambda$ and $\lambda \in \Lambda$; then $\pi: \Lambda / \Lambda' \to \Lambda / \Lambda'$ defined as $\pi(v) = (v+\lambda) \Mod \Lambda'$
%\begin{align*}
%  \pi \colon \Lambda / \Lambda' &\to \Lambda %/ \Lambda'
%  ; \; \pi(v) = (v+\lambda) \Mod \Lambda'
%\end{align*}
is a permutation of $\Lambda / \Lambda'$.
\end{lem}
%%%% now a probability tools
%%%%%%%%%%%%%%%%%%%%%%%%%
%%%%%%%%%%%%%%%%%%%%%%%%%
%%%%%%%%%%%%%%%%%%%%%%%%
\subsubsection*{Cyclotomic fields and modules}
For a power of $2$ integer $m \geq 1$, the $m^{\text{th}}$ \emph{cyclotomic number field} is the extension $K=\mathbb{Q}(\zeta_m)$ with degree $n=m/2$, where $\zeta_m$ is any $m^{\text{th}}$ primitive root of unity. The \emph{ring of integers} of $K$ is $R= \mathbb{Z}[\zeta_m] \cong \mathbb{Z}[X]/\left(X^n+1 \right)$ and given $q>0$, its quotient ring is $R_q=\mathbb{Z}_q[X]/\left(X^n+1 \right)$. For an integer $d>0$, a subset $M \subseteq K^d$ is an $R$-module if it is closed under addition and multiplication by elements of $R$.
\subsubsection*{Polynomial splitting}
We will use a polynomial splitting in section \ref{proba} to analyse our error probability bound, similarly to \cite[Section C]{NEWHOPE}. For that, we write $n=n_0 \times L$ and take $S=\mathbb{Z}[Y]/\left(Y^{n_0}+1\right)$.\\
%with $n_0=8$ and $L=32$. 
%Recall that $R=\mathbb{Z}[X]/\left(X^n+1\right)$ 
Given a polynomial $a(X) \in R$, we define for $\kappa=0, \dots , L-1$ the vectors $a^{(\kappa)}$ as 
$a^{(\kappa)}=(a_{\kappa}, a_{\kappa+L}, a_{\kappa+2L}, \dots , a_{\kappa+n-L})$.
Taking $Y=X^L$ allows to write $a(X)= \sum_{\kappa=0}^{L-1}a^{(\kappa)}(Y)X^{\kappa}$. This in turn allows to express the multiplication $a(X)b(X)$ as $p(X):=a(X)b(X)=\sum_{\kappa=0}^{L-1}p^{(\kappa)}(Y)X^{\kappa}$. Hence,
$$p^{(\kappa)}(Y)=\sum_{i=0}^{L-1}  Y^{\delta_{i,\kappa}} \, a^{(i)}(Y) \cdot b^{(\kappa-i \Mod L)}(Y),$$
where $\delta_{i,\kappa}$ is either $0$ or $1$. In the sequel we take out the $\Mod L$ operation to simplify notations. 
\subsection{$E_8$ lattice} \label{E8}
The $E_8$ lattice is a discrete subgroup of $\mathbb{R}^8$ of full rank \cite[p.121]{Conway}. One possible basis for $E_8$ is given by the rows of the matrix 
$$ \mathbf{E}=
\begin{bmatrix}
\text{\small $2$} & \text{\small $0$} &\text{\small $0$}&\text{\small $0$}&\text{\small $0$}&\text{\small $0$}&\text{\small $0$}&\text{\small $0$} \\
\text{\small $-1$} & \text{\small $1$} &\text{\small $0$}&\text{\small $0$}&\text{\small $0$}&\text{\small $0$}&\text{\small $0$}&\text{\small $0$} \\
\text{\small $0$} & \text{\small $-1$} &\text{\small $1$}&\text{\small $0$}&\text{\small $0$}&\text{\small $0$}&\text{\small $0$}&\text{\small $0$} \\
\text{\small $0$} & \text{\small $0$} &\text{\small $-1$}&\text{\small $1$}&\text{\small $0$}&\text{\small $0$}&\text{\small $0$}&\text{\small $0$} \\
\text{\small $0$} & \text{\small $0$} &\text{\small $0$}&\text{\small $-1$}&\text{\small $1$}&\text{\small $0$}&\text{\small $0$}&\text{\small $0$} \\
\text{\small $0$} & \text{\small $0$} &\text{\small $0$}&\text{\small $0$}&\text{\small $-1$}&\text{\small $1$}&\text{\small $0$}&\text{\small $0$} \\
\text{\small $0$} & \text{\small $0$} &\text{\small $0$}&\text{\small $0$}&\text{\small $0$}&\text{\small $-1$}&\text{\small $1$}&\text{\small $0$} \\
\text{\scriptsize $1/2$} & \text{\scriptsize $1/2$} &\text{\scriptsize $1/2$}&\text{\scriptsize $1/2$}&\text{\scriptsize $1/2$}&\text{\scriptsize $1/2$}&\text{\scriptsize $1/2$}&\text{\scriptsize $1/2$}
\end{bmatrix}
$$
This lattice has $2$ types of Voronoi-relevant vectors of the form $ (\pm 1^2, 0^{6})\in \text{VR}_1$ and  $(\pm 0.5^8)\in \text{VR}_2$. Note that $| \text{VR}_1 |=112$ and $| \text{VR}_2 |=128$, so that the total is $240$. The volume of $E_8$ is simply $1$. A simple and fast CVP algorithm for $E_8$ is given in \cite{Fastcvp}. Note that $2 \mathbb{Z}^8 \subset E_8 \subset \tfrac{1}{2}\mathbb{Z}^8$.
%%%%%%%%%%%%%%%%%%%%%%%%%%%%%
\subsection{Gaussian-like Error Distribution}
When dealing with module-LWE defined below, we work with a Gaussian-like error distribution over the ring $R^d$.

Since it is challenging to implement a discrete Gaussian sampler which is efficient and protected against timing attacks, one can replace the secret and error distribution by the centered binomial distribution $\psi_k$ of standard deviation $\sqrt{k/2}$ introduced in \cite{NEWHOPE}, which is defined as
$$\psi_k=\sum_{i=1}^{k}(b_i-b'_i)$$ 
where $b_i,b'_i$ are independent and uniformly distributed in $\{0,1\}$, for $i=1, \dots ,k$.

Note that choosing $\psi_k$ as error distribution does not significantly decrease security compared to a rounded Gaussian distribution, and this can be  shown with a Rényi divergence-based analysis, as in {\cite[Theorem 4.1]{NEWHOPE}} for $k=16$. It remains true for a general value of $k$ {\cite[Section 5.3]{analysis}}.

If $x \in R$, we write $x \leftarrow \Psi_k$ to mean that
$x \in R$ is generated from a distribution where each of its
coefficients is generated according to $\psi_k$.
Similarly, a $d$-dimensional vector $\mathbf{x} \in R^d$ can be generated
according to the distribution $\Psi_k^d$.
%We define the Gaussian distribution $D_{r'}$ with parameter $r'$ over $K\otimes_{\mathbb{Q}} \mathbb{R}$ to output an element $a \in K\otimes_{\mathbb{Q}} \mathbb{R}$ for which $\sigma(a) \in H$ is an $N$-dimensional i.i.d. Gaussian distribution with zero mean and covariance $r'^2$.
%In our application, we use error distributions of the form $\psi= N D_{r'}$ with parameter $N r'$ and discretize it to $R$, by discretizing first $D_{r'}$ to $R^{\vee}$ using coordinate-wise randomized rounding \citep{TOOLKIT}, and denote the resulting distribution by $\chi= \lfloor \psi \rceil_{R}$. Note that $\chi$ can be seen as a subgaussian of parameter $r=\mathcal{O}(N r')$ \cite[Lemma 8.2]{TOOLKIT}.
%%%%%%%%%%%%%%%%%%%%%%%%%%%%%%%%%%%%%%%
%%%%%%%%%%%%%%%%%%%%%%%%%%%%%
%\begin{prop}[\cite{TOOLKIT}, Lemma 8.2] \label{rounded_subgaussian} \normalfont
%If $D_{r'}$ is a continuous Gaussian with parameter $r' \geq 1$, and we use coordinate-wise randomized rounding, then $\chi= \lfloor \psi \rceil_{R}$ is subgaussian with parameter $r=N \sqrt{r'^2+2 \pi / N}=\mathcal{O}(N r')$.
%\end{prop}
%%%%%%%%%%%%%%%%%%%%%%%
%%%%%%%%%%%%%%%%%%%%%%%%%%%%%%%%
% One can have a better parameter from Lemma 50 in \cite{RLWE_DIST}, where $r$ becomes $\sqrt{r'^2+1/4 \cdot \text{rad}(m)/m}=\mathcal{O}(r').$
%\subsubsection*{Cryptographic definitions} 
\subsection{Cryptography and Ring-LWE} 
\subsubsection*{Cryptographic definitions} 
%A function is \emph{negligible} in $N$ if it is asymptotically smaller than $N^{-c}$ for any constant $c > 0$.
%Two families of distributions are \emph{computationally indistinguishable} if the probability of telling the difference between them using any efficient distinguisher algorithms $\mathcal{D}$ is negligible in $N$.\\
We define the notion of \emph{key encapsulation mechanism} (KEM) following \citep{PEIKERT}, which consists of three algorithms $(\textsf{Gen}, \textsf{Encaps}, \textsf{Decaps})$, where \textsf{Gen} takes a public parameter $pp$ and returns a secret key $sk$ and a private one $pk$, where \textsf{Encaps} takes $(pp, pk)$ to produce a ciphertext $c$ and a key $\mathbf{k} \in \mathcal{K}$, and where \textsf{Decaps} takes the secret key $sk$ and ciphertext $c$ to return a key $\mathbf{k} \in \mathcal{K}$ or the symbol $\perp$ to denote rejection. A KEM satisfies IND-CPA security, if the outputs of the following
\enquote{real} and \enquote{ideal} games are computationally indistinguishable:
\begin{center}
\footnotesize
\begin{tabular}{ c  | c }
  \textbf{Real Game}  & \textbf{Ideal Game} \\
 $ (pk,sk) \leftarrow \textsf{Gen}(pp)$ & $ (pk,sk) \leftarrow \textsf{Gen}(pp)$ \\
 $(c,\mathbf{k}) \leftarrow \textsf{Encaps}(pp,pk)$ & $(c,\mathbf{k}) \leftarrow \textsf{Encaps}(pp,pk)$ \\
  & $ \mathbf{k}^* \leftarrow \mathcal{K}$ \\
$\text{Output}(pp,pk,c,\mathbf{k})$ & $\text{Output}(pp,pk,c,\mathbf{k}^*)$
\end{tabular}
\end{center}
%%%%%%%%%%%%%%%%%%%%%%%%%%%
\subsubsection*{Module-LWE ($M$-LWE)} The security of our scheme is based on the hard \emph{module-LWE} problem \cite{module-lwe}. Let $d$ be a positive integer parameter. The problem consists in distinguishing uniform samples $(\mathbf{a}_i,b_i) \leftarrow R_q^d \times R_q$ from samples $(\mathbf{a}_i,b_i) \leftarrow R_q^d \times R_q$ where $\mathbf{a}_i \leftarrow R_q^d$ is uniform and $b_i = \mathbf{a}_i \cdot \mathbf{s}+e_i$ with $\mathbf{s} \leftarrow \Psi_k^d$ common to all samples and $e_i \leftarrow \Psi_k$ fresh for every sample. The multiplication $\mathbf{a}_i \cdot  \mathbf{s}$  is a dot product $a_1 \cdot s_1 + \dots + a_d \cdot s_d$, such that each $a_i \cdot s_i$ is a polynomial product modulo $(X^n+1)$.
\section{Key generation algorithm} \label{alg}
We give here the key generation algorithm between Alice and Bob.
\begin{table}[htb]
\centering
\begin{tabular}{|p{3.3cm} p{1cm} p{3.2cm}|} 
\hline
\multicolumn{3}{|c|}{Parameters: $q=2^{11}$; $k=2$; $n =256$; $d=3$} \\ [0.5ex] 
\hline\hline
\textbf{Alice (server)} &  & \textbf{Bob (Client)} \\ 
$\mathbf{A} \xleftarrow{\text{\$}} R_q^{d \times d} $ &  &   \\
$\mathbf{s}, \mathbf{e} \xleftarrow{\text{}} \Psi_k^d$ &  & $\mathbf{s'}, \mathbf{e'} \xleftarrow{\text{}} \Psi_k^d$, $e''  \xleftarrow{\text{}} \Psi_k$  \\
$\mathbf{b} := \mathbf{A} \mathbf{s}+\mathbf{e} \in R_q^d$ & $\xrightarrow{(\mathbf{A},\mathbf{b})}$ & \\ 
 & & $\mathbf{u} := \mathbf{A}^T\mathbf{s'}+\mathbf{e'} \in R_q^d$ \\
 & & $v := \mathbf{b} \cdot \mathbf{s'}+e'' \in R_q$\\
 & $\xleftarrow{(\mathbf{u},r)}$  & $r= \text{HelpRec}(v)$ \\
 $v' := \mathbf{u} \cdot \mathbf{s} \in R_q$ & & \\
  $ \hat{\mathbf{k}} = \text{Rec}(v',r) $ & & $ \mathbf{k} = \text{Rec}(v,r)$  \\[1ex] 
\hline
\end{tabular}
\vspace{.1mm}
\caption{\normalfont Key generation algorithm based on module-LWE \label{table1}}
% Put the label inside to get a correct labeling}
\end{table}
Our protocol makes use of the following  lattices of dimension $n=256$: the quantization lattice $\Lambda_1= \left( \frac{q}{2^p} E_8 \right)^{32}$ for some integer $p \geq 1$, the coding lattice $\Lambda_2 = \left( \frac{q}{2}  E_8 \right)^{32}$ and the shaping lattice $\Lambda_3=q \left( \mathbb{Z}^8 \right)^{32}$.
This choice implies that $\Lambda_3 \subseteq \Lambda_2 \subseteq \Lambda_1$.
The \emph{key rate} given by $R_K=\frac{1}{n} \log_2 \left(\frac{\text{Vol}(\Lambda_3)}{\text{Vol}(\Lambda_2)}\right)$ is simply $1$ so that the protocol provides $256$ bits of key. Furthermore, the reconciliation rate $R_P=\frac{1}{n} \log_2 \left(\frac{\text{Vol}(\Lambda_2)}{\text{Vol}(\Lambda_1)}\right)$ is calculated to be $p-1$.

With regard to Table \ref{table1}, the KEM algorithm consists of taking a random matrix $\mathbf{A}$ from $R_q^{d \times d}$ by Alice, and referring it as the public parameter $pp$. Then she chooses $\mathbf{e}, \mathbf{s} \leftarrow \Psi_k^d$, computes $\mathbf{b}= \mathbf{A}\mathbf{s}+\mathbf{e}$, and outputs a public key $pk=\mathbf{b}$ and a secret key $sk=\mathbf{s}$. When it's Bob's turn, he chooses independent $\mathbf{e'}, \mathbf{s'} \leftarrow \Psi_k^d$ and $e'' \leftarrow \Psi_k$, then computes $\mathbf{u} = \mathbf{A}^T\mathbf{s'}+\mathbf{e'} \in R_q^d$ and  $v = \mathbf{b} \cdot \mathbf{s'}+e'' \in R_q$. He outputs $c=\left( \mathbf{u}, r \right) \in R_q^d \times \Lambda_1/\Lambda_2$ with 
$r=\text{HelpRec}(v):= Q_{\Lambda_1}(v) \Mod \Lambda_2$ and $\mathbf{k}$ in $\Lambda_2 / \Lambda_3$ such that $\mathbf{k}= \text{Rec}(v,r):=Q_{\Lambda_2}\left(v-r\right) \Mod \Lambda_3$.
At the end, Alice computes $v'=\mathbf{u} \cdot \mathbf{s}$, and outputs $\hat{\mathbf{k}}=\text{Rec}(v',r)$. 

Note that when $v \in R_q$, the notation $Q_{\Lambda_1}(v)$ means that we perform $Q_{\frac{q}{2^p} E_8}$ on each component $v^{(\kappa)}$ (see Section \ref{preliminaries}) where $n_0=8$ and $L=32$, and similarly for $Q_{\Lambda_2}(\cdot)$, $\Mod \Lambda_2$ and $\Mod \Lambda_3$ operations.
%%%%%%%%%%%%%%
\begin{rem}[Comparison with NewHope and Kyber \cite{NEWHOPE, kyber, kyber3})]
Note that this algorithm is reconciliation-based, similarly to \cite{PEIKERT} and the first version of the NewHope protocol \cite{NEWHOPE}. 
For instance, in \cite{NEWHOPE} the functions HelpRec and Rec can be written as the above form by taking the product lattices $\Lambda_1=(q \tilde{D}_4/2^p)^{256}$, $\Lambda_2=(q \tilde{D}_4)^{256}$ and $\Lambda_3=q \mathbb{Z}^{1024}$.
We point out that a dither is not required in our algorithm like in \citep{PEIKERT,NEWHOPE} since the modulus $q$ is an even number.
Unlike NewHope, the proposed protocol is based on Module-LWE and uses the same parameters $n, d, k$ as in CRYSTALS - Kyber \cite{kyber,kyber3}. In order to obtain $256$ bits of key with $n=256$, a higher-dimensional lattice than $\tilde{D}_4$ is needed.
\end{rem}
%\paragraph{Construction using Barnes-Wall lattices}\label{para}
%For an explicit construction we choose $\Lambda_1= \Lambda_0$ and $\Lambda_2= T \cdot BW^{n-1}$, where $T=\beta \sqrt{N} \cdot \Phi$ and $\beta $ a power of $2$. By this choice, all the operations with $\Lambda_2$ in Table \ref{table:kysymys} can be deduced from Section \ref{E8}.
%The operation $\textsc{ParBW}_T$ corresponds to a quantization operation $Q_{\Lambda_2,\mathcal{P}}$ induced by a partition $\mathcal{P}$ of the Barnes-Wall lattice: $\textsc{ParBW}_T = Q_{\Lambda_2,\mathcal{P}}$ (see Theorem \ref{linearity}).
%Since $ \beta BW^{n-1} \subseteq BW^{n-1} \subseteq \mathbb{Z}^N$, we obtain that $ \Lambda_2 \subseteq \Lambda_1 = \Lambda_0.$
%For the inclusion $\Lambda_3 \subseteq \Lambda_2$, we must have $q \mathbb{Z}^N \subseteq \beta BW^{n-1}$, or $\frac{q}{\beta} \mathbb{Z}^N \subseteq  BW^{n-1}$. By Proposition \ref{main_prop}, this is true when
%\begin{equation} \label{beta_cond}
%q / \beta = 2^k, \text{with } k \geq \floor{n/2}.
%\end{equation}
\section{Error probability bounds} \label{proba}
In this section, we provide more technical details on estimating the error probability $P_e = \mathbb{P} \{ \mathbf{k} \neq \hat{\mathbf{k}} \}$. We will prove that in our described protocol, the parameter set we recommend in Table \ref{table1} yields $P_e< 2^{-174}$. 
%%%%%%%%%%%%%%%%%%%

%%%%%%%%%%%%%%%%%%%%%%%%%%%%%%%%%%%
\subsection{Reliability condition}
According to Section \ref{alg}, the two keys $\mathbf{k}$ and $\hat{\mathbf{k}} $ would be identical whenever $Q_{\Lambda_2}\left(v-r \right) \Mod \Lambda_3$ and $Q_{\Lambda_2}\left(v'-r \right) \Mod \Lambda_3$ are equal. Setting $e_Q=v-Q_{\Lambda_1}(v)$ we can show that a sufficient condition is  $Q_{\Lambda_2}\left( (v-v')+e_Q \right)=0$; or more appropriately:
$$Q_{ \frac{q}{2}\left(1-\frac{1}{2^{p-1}}\right) E_8}\left( v-v' \right)=0, $$
because $e_Q \in \mathcal{V}(\Lambda_1)$. For clearer notations, we assign $C:=\frac{q}{2}\left(1-\frac{1}{2^{p-1}}\right)$.
Let $\omega$ denote the error difference between $v$ and $v'$ for which
\begin{flalign*}
v-v' & = \mathbf{b} \cdot \mathbf{s'}+e''-\mathbf{u} \cdot \mathbf{s} && \\ \nonumber
&= \left( \mathbf{A}\mathbf{s}+ \mathbf{e} \right) \cdot \mathbf{s'} +e'' - \left( \mathbf{A}^T \mathbf{s'}+ \mathbf{e'} \right) \cdot \mathbf{s} && \\ \nonumber
&= \mathbf{e} \cdot \mathbf{s'} - \mathbf{e'} \cdot \mathbf{s}+ e'' \in R_q && \\ \nonumber
&= {e_1s'_1 + \dots + e_ds'_d} +  {-e'_1s_1 - \dots - e'_ds_d} + e'' && \\ \nonumber
&= \omega_1 + \omega_2 + \dots \omega_d + e'', \text{ where } \omega_i = e_is'_i-e'_is_i
\end{flalign*}
This can be written in polynomial form as:
$$ \omega(X)= \omega_1(X) + \omega_2(X) + \dots \omega_d(X) + e''(X)$$
So for ${\kappa}=0, \dots, L-1$, the expression of $\omega^{(\kappa)}(Y)$ will be:
$$ \sum_{j=1}^d \sum_{i=0}^{L-1} Y^{\delta_{i,{\kappa}}} \left[ {e_j}^{(i)}  {s'}_j^{({\kappa}-i)}(Y) - {e'}_j^{(i)}  s_j^{({\kappa}-i)}(Y)\right] + {e''}^{(\kappa)}(Y).$$
%%%%%%%%%%%%%%%%%%%%%%%%%%%%%%%%%%ù
As in \cite{NEWHOPE, vanpop}, we will consider a union bound over all Voronoi-relevant vectors.  Note that $\omega^{(\kappa)}(Y)$ is still a polynomial with $8$ coefficients. Decoding will be correct if $ \omega^{(\kappa)} \in C \cdot \mathcal{V}(E_8)$ for all ${\kappa}=0,1,2, \dots , L-1$. More formally,
$ \omega^{(\kappa)} \in C \cdot \mathcal{V}(E_8) \Longleftrightarrow \langle \omega^{(\kappa)} , v \rangle \leq \frac{\| v\|_2^2}{2}, \forall v \in C(\text{VR}_1 \cup \text{VR}_2)$.\\
We mention that multiplying the vector form of $e^{(i)}(Y)$ by $Y$ is equivalent to a right shift with a minus sign on the first term.
%So when an expression of the form $e^{(i)}(Y)  s'^{({\kappa}-i)}(Y)$ is multiplied by $Y$, one can choose the vector to shift. 
Using this and the fact that the distributions of $e_j^{({\kappa}-i)}$ and ${e'_j}^{({\kappa}-i)}$ are invariant
by conj$(\cdot)$ and by multiplication by $-1$, as well as the distributions of ${s_j}^{(i)}$ and ${s'_j}^{(i)}$ are also invariant by shifting and by multiplication by $-1$, we obtain a more compact form of $\langle \omega^{(\kappa)} , v \rangle$:
$$ \langle \omega^{(\kappa)} , v \rangle = \left \langle (\tilde{s'}_j,\tilde{s}_j)_{j=1, \dots, d} , W_{v,{\kappa}} \right\rangle + \langle e''^{(\kappa)}  , v \rangle$$
where $\tilde{s}_j$ and $\tilde{s}'_j$ are $n$ dimensional vectors of independent centered binomial coefficients, and 
$$W_{v,{\kappa}}= C \cdot \left[  \, \text{conj}\left( e_j^{({\kappa}-i)} \right) \cdot v , \dots , \text{conj}\left( {e'_j}^{({\kappa}-i)} \right) \cdot v \right]_{\substack{i=0, \dots, L-1, \\ j =  1,\dots,d } }$$
can be identified to
$W_{v,{\kappa}}= C \cdot \left( e^{(0)}\cdot v , \dots e^{(2L \times d-1)}\cdot v \right)$, where each component $e^{(i)} \cdot v$ is a polynomial multiplication of an $8$-dimensional vector $e^{(i)}$ by a Voronoi-relevant vector $v$. The multiplication is done modulo $(Y^{8}+1)$. Moreover, the $e^{(i)}$s are independent with centered binomial coefficients, distributed also independently. For instance, if $v_1 \in \text{VR}_1$ is a Voronoi-relevant vector of type 1, then $W_{v_1,{\kappa}}$ is given by the general form
$$\tfrac{C}{2} \cdot \left[ (\pm  e^{(0)}_{i_0} \pm e^{(0)}_{i_1} , \dots ,\pm  e^{(0)}_{i_6} \pm  e^{(0)}_{i_7}) ; \dots \right].$$ 
%%%
However, if $v_2 \in \text{VR}_2$ is a Voronoi-relevant vector of type 2, then each component of $e^{(i)} \cdot v$ of $W_{v_2,{\kappa}}$ is of the form:
$$\tfrac{C}{4} (\pm  e^{(i)}_{i_0} \pm  e^{(i)}_{i_1} \pm  e^{(i)}_{i_2} \pm  e^{(i)}_{i_3} \pm  e^{(i)}_{i_4} \pm  e^{(i)}_{i_5} \pm  e^{(i)}_{i_6} \pm  e^{(i)}_{i_7}).$$
\subsection{Error probability calculations}
Recall that an error occurs if $\omega^{(\kappa)} \notin C \cdot \mathcal{V}(E_8)$ for some ${\kappa}= 0, \dots ,  L-1$. So one can bound $P_e$ by
$$
 \mathbb{P}\left\{ \exists {\kappa}, \; \exists v \in C (\text{VR}_1 \cup \text{VR}_2)  \, : \,  \langle \omega^{(\kappa)} , v \rangle > \frac{\| v\|_2^2}{2} \right\}
$$
Using the fact that 
$\langle \omega^{(\kappa)} , v \rangle = \langle (\tilde{s'},\tilde{s}) , W_{v,{\kappa}} \rangle + \langle e''^{(\kappa)}  , v \rangle$ we obtain:
\begin{flalign*}
P_e 
& \leq \sum_{{\kappa}=0}^{L-1} \mathbb{P}\left\{ \exists v  \, : \, \langle (\tilde{s'},\tilde{s}) , W_{v,{\kappa}} \rangle > \frac{\| v\|_2^2}{2} -  \langle e''^{(\kappa)}  , v \rangle   \right\} 
\end{flalign*}
Observing that $\frac{||v||_2^2}{2} - \langle e''^{(\kappa)},v \rangle \geq C^2-2kC$ for type 1 vectors (resp. $C^2-4kC$ for type 2), we can bound each term by computing the distribution of $ \langle (s',s),W_{v,{\kappa}} \rangle$, which is a sum of $192$ i.i.d. random variables of the form $e^{(i)} \cdot v$. Details are omitted due to lack of space. From our numerical simulations, we obtain the following table:
\begin{table}[htb] %To appear belowe text
    \begin{center}
\begin{tabular}{ | p{3cm} | p{1cm}  | p{1cm} | p{1cm}| p{1cm}|}
 \hline
 \textbf{Error Probability Bound} & $p=2$ & $p=3$ & $p=4$ & $p=5$ \\
 \hline
  $q=2^{11}$, $k=2$  & $2^{-48}$ & $2^{-113}$  & $2^{-153}$ & $2^{-174}$ \\
\hline
 $q=2^{12}$, $k=4$   & $2^{-47}$ & $2^{-112}$   &  $2^{-152}$ & $2^{-172}$ \\
\hline
% $q=7681$  & $2^{-171}$ & $2^{-350}$  & $2^{-451}$  \\
%\hline
 $q=2^{13}$, $k=4$  & $2^{-193}$ & $2^{-390}$  & $2^{-499}$ & $2^{-557}$ \\
\hline
\end{tabular}
\end{center}
\caption{ \normalfont Upper bound for error probability for different values of moduli $q$, noise parameter $k$ and reconciliation rate parameter $p$ \label{tableproba}}
\end{table}
%The modulus to noise ratio $q/r$ defined in \cite{PEIKERT} in our scheme is of order $\tilde{O}\left(N^{7/4}\right)$, i.e. the same as in \cite{PEIKERT}.
%\begin{rem}
%The values of $q$ in equation (\ref{condi2}) are chosen to guarantee IND-CPA security assuming the hardness of $R$-LWE (see Section \ref{security}). In practice, many proposed $R$-LWE-based protocols use smaller $q$ and estimate practical security against known classes of attacks. See discussion in Section \ref{cbd}.
%\end{rem}
%In \cite[Section 4.4]{PEIKERT}, $q/r$ was shown to be of order $\tilde{O}\left( N^{7/4} \right) $. Unfortunately, we can't do better than $7/4$ on the exponent using our strategy.
%\begin{rem}
%Note that no \vv{folding} mod $q$ of the coefficients of $v'$ occurs within the decoding radius of the bounded distance decoder. In fact, the ball of radius $d_{\text{min}}(\Lambda_2)/2$ is contained in the square of side $q/(2 \sqrt{N})$ in the canonical embedding. This guarantees correct decoding within that radius.
%\end{rem}
%%%%%%%%%%%%%%%%%%%%%%%%%%%%%%%%%
%%%%%%%%%%%%%%%%%%%%%%%%%%%%%%%%%
\section{IND-CPA Security} \label{security}
We will prove that, with the choice of $q$ in Section \ref{proba}, the algorithm is \textit{IND-CPA} secure, assuming the hardness of $\text{$M$-LWE}$ given two samples. This proof is generic and holds in the setting of the key generation protocol in Section \ref{alg} independently of the choice of the lattices $\Lambda_1$ and $\Lambda_2$ as long as the CVP can be done efficiently.
We follow the same argument as Section 4.2 in \cite{PEIKERT}. We consider the adjacent games below:
\begin{center}
\footnotesize
\begin{tabular}{ C{4.4cm} | c }
  \textbf{Game 1}  & \textbf{Game 1 '} \\
 $\mathbf{A}\xleftarrow{\text{\$}} R_q^{d \times d}$ &  $\mathbf{A}\xleftarrow{\text{\$}} R_q^{d \times d}$ \\ 
 $ (\mathbf{b},\mathbf{s}) \leftarrow \textsf{Gen}(\mathbf{A}) $ & $ (\mathbf{b},\mathbf{s}) \leftarrow \textsf{Gen}(\mathbf{A})$ \\
 $\left((\mathbf{u},r),\mathbf{k}\right) \leftarrow \textsf{Encaps}(\mathbf{A},\mathbf{b})$ & $\left((\mathbf{u},r),\mathbf{k}\right) \leftarrow \textsf{Encaps}(\mathbf{A},\mathbf{b})$ \\
  & $ \mathbf{k}^* \xleftarrow{\text{\$}} \Lambda_2/\Lambda_3$ \\
$\text{Output}\left(\mathbf{A},\mathbf{b},(\mathbf{u},r),\mathbf{k}\right)$ & $\text{Output}\left(\mathbf{A},\mathbf{b},(\mathbf{u},r),\mathbf{k}^*\right)$
\end{tabular}
\end{center}
\begin{center}
\footnotesize
%%%%%%%%%%%%%%%%%%%%%%%%%%%%%%%%%%%%
\begin{tabular}{ C{3.9cm} | c }
  \textbf{Game 2}  & \textbf{Game 3} \\
 $\mathbf{A}\xleftarrow{\text{\$}} R_q^{d \times d}$ &  $(\mathbf{A},\mathbf{b})\xleftarrow{\text{\$}} R_q^{d \times d}\times R_q^d$ \\ 
 $ \mathbf{b}\xleftarrow{\text{\$}} R_q^d $ & $ (\mathbf{u},v)\xleftarrow{\text{\$}} R_q^d\times R_q $ \\
 $\left((\mathbf{u},r),\mathbf{k}\right) \leftarrow \textsf{Encaps}(\mathbf{A},\mathbf{b})$ & $r=\text{HelpRec}(v)$ \\
  & $ \mathbf{k}^* \xleftarrow{\text{\$}} \Lambda_2/\Lambda_3$ \\
$\text{Output}\left(\mathbf{A},\mathbf{b},(\mathbf{u},r),\mathbf{k}\right)$ & $\text{Output}\left(\mathbf{A},\mathbf{b},(\mathbf{u},r),\mathbf{k}^*\right)$
\end{tabular}
\end{center}

Notice that Game 1 is the \enquote{real} game defined in Section \ref{preliminaries}, and Game 1' is the \enquote{ideal} one. Our aim is to prove that Game 1 and Game 1' are computationally indistinguishable. We'll do so sequentially.

%To do that, we will prove the indistinguishability between Game 1 and Game 2, then Game 2 and Game 3, and at the end between Game 3 and Game 1'.

Clearly Game 1 and Game 2 are computationally indistinguishable under the assumption of hardness of $\text{$M$-LWE}$.

To prove that Game 2 and Game 3 are computationally indistinguishable, we use the following Theorem which is essentially a consequence of the Crypto Lemma \cite[Lemma 4.1.1]{ZAMIR}. It guarantees uniformity of the key without a dither. 
\begin{ther} \label{claim peikert} \normalfont
If $v \in R_q$ is uniformly random, then $\mathbf{k}=\text{Rec}(v,r)$ is uniformly random, given $r=\text{HelpRec}(v)$.
\end{ther}
\begin{IEEEproof}
For fixed $\mathbf{k}, \mathbf{k}' \in \Lambda_2 / \Lambda_3$, we define $\forall \, v \in R_q $,
%\vspace{-2mm}
$$\pi_{\mathbf{k}, \mathbf{k}'}(v)=\left(v-\mathbf{k} +\mathbf{k}'\right) \Mod \Lambda_3.$$
%\begin{align*}
%  \pi_{\mathbf{k}, \mathbf{k}'} \colon \Lambda_2 / \Lambda_3 &\to \Lambda_2 / \Lambda_3\\
%  v &\mapsto \pi_{\mathbf{k}, \mathbf{k}'}(v)=\left(v-\mathbf{k} +\mathbf{k}'\right) \Mod \Lambda_3 .  
%\end{align*}
Notice that $ \pi_{\mathbf{k}, \mathbf{k}'}(v) \in R_q$ because $(-\mathbf{k}+\mathbf{k}') \in \Lambda_2 \subseteq \mathbb{Z}^n$ and hence $\pi_{\mathbf{k}, \mathbf{k}'}(v) \in \mathbb{Z}_q^n$. So $\pi_{\mathbf{k}, \mathbf{k}'}$ is a permutation of $R_q$ by Lemma \ref{perm lemma}. The proof of Theorem \ref{claim peikert} results from these lemmas:
\begin{lem}\label{(first_lemma)} \normalfont
$\forall \; \mathbf{k}, \mathbf{k}' \in \Lambda_2 / \Lambda_3$ and $\forall \; v \in R_q$ we have 
$ \text{HelpRec}(v)=\text{HelpRec}\left(\pi_{\mathbf{k}, \mathbf{k}'}(v)\right)$.
\end{lem}
\begin{IEEEproof}
\vspace{-1mm}
\begin{align*}
r' &= Q_{\Lambda_1}\left(\pi_{\mathbf{k}, \mathbf{k}'}(v)\right) \Mod \Lambda_2 &&\\ \nonumber
 		&= Q_{\Lambda_1}\left(\left(v-\mathbf{k}+ \mathbf{k}'\right) \Mod \Lambda_3\right) \Mod \Lambda_2 &&\\ \nonumber 		
 		&= Q_{\Lambda_1}\left(v-\mathbf{k}+ \mathbf{k}'- Q_{\Lambda_3}\left(v-\mathbf{k}+ \mathbf{k}'\right)\right) \Mod \Lambda_2 &&\\ \nonumber
 		&= \left(Q_{\Lambda_1}\left(v\right)-\mathbf{k}+ \mathbf{k}'- Q_{\Lambda_3}\left(v-\mathbf{k}+ \mathbf{k}'\right) \right) \Mod \Lambda_2  &&\\ \nonumber
 		&= Q_{\Lambda_1}\left(v\right)\Mod \Lambda_2 = r. \tag*{\IEEEQED} 
\end{align*} 
\let\IEEEQED\relax%
\end{IEEEproof}
\vspace{-3mm}
\begin{lem}\label{(second_lemma)} \normalfont
Suppose that $\mathbf{k}$ $=\text{Rec}(v,r)$ $=Q_{\Lambda_2}\left(v-r\right)$ $ \Mod \Lambda_3$. Then 
$\forall \; \mathbf{k}' \in \Lambda_2 / \Lambda_3$ we have 
$ \mathbf{k}'= \text{Rec}(\pi_{\mathbf{k},\mathbf{k}'}(v),r).$
\end{lem}
\begin{IEEEproof}
{\allowdisplaybreaks
\begin{align*}
\text{Rec}(\pi_{\mathbf{k},\mathbf{k}'}(v),r)
		%&= Q_{\Lambda_2}\left(v-\mathbf{k}+ \mathbf{k}'- Q_{\Lambda_3}\left(v-\mathbf{k}+ \mathbf{k}'\right)-r\right) \Mod \Lambda_3 &&\\ \nonumber
		%&= \left[ Q_{\Lambda_2}\left(v-r\right) -\mathbf{k}+ \mathbf{k}'- Q_{\Lambda_3}\left(v-\mathbf{k}+ \mathbf{k}'\right) \right]\Mod \Lambda_3 &&\\ \nonumber
		&= \left[ Q_{\Lambda_2}\left(v-r\right) -\mathbf{k}+ \mathbf{k}' \right]\Mod \Lambda_3 &&\\ \nonumber
		&= \left[ Q_{\Lambda_2}\left(v-r\right) \Mod \Lambda_3 -\mathbf{k}+ \mathbf{k}' \right]\Mod \Lambda_3 &&\\ \nonumber
		&	= \left[ \mathbf{k} -\mathbf{k}+ \mathbf{k}' \right]\Mod \Lambda_3 = \mathbf{k}'.
\tag*{\IEEEQED} 
\end{align*} 
}
\let\IEEEQED\relax%
\end{IEEEproof}
%Hence we have proved the following:
\vspace{-4mm}
\begin{corol}\label{(111)} \normalfont
$\forall \; \mathbf{k}, \mathbf{k}' \in \Lambda_2 / \Lambda_3$ and $\forall \; v \in R_q$, there exist $ v'=\pi_{\mathbf{k},\mathbf{k}'}(v)$ such that
$ \text{HelpRec}(v)=\text{HelpRec}\left(\pi_{\mathbf{k}, \mathbf{k}'}(v) \right), $ and
$ \mathbf{k}= \text{Rec}(v,r) \Longleftrightarrow\mathbf{k}'= \text{Rec}(\pi_{\mathbf{k},\mathbf{k}'}(v),r).$
\end{corol}
We conclude the proof of Theorem \ref{claim peikert} by showing that $\mathbf{k}$ is uniform and independent of $r$ when $v$ is uniform:
\begin{flalign*}
\mathbb{P}\{ \mathbf{k} \mid r \}
		&= \sum_{v\in R_q} \mathbb{P}\{v\} \cdot \mathbb{P}\{ \mathbf{k} \mid r, v \} &&\\ \nonumber
		&=  \sum_{v\in R_q} \mathbbm{1}_{ \left\{ \substack{r=\text{HelpRec}(v)\\ \mathbf{k}=\text{Rec}(v,r)} \right\} } \cdot \mathbb{P}\{v\} &&\\ \nonumber
 		&
 		=  \sum_{v\in R_q} \mathbbm{1}_{ \left\{ \substack{r=\text{HelpRec}(\pi_{\mathbf{k},\mathbf{k}'}(v))\\ \mathbf{k}'=\text{Rec}(\pi_{\mathbf{k},\mathbf{k}'}(v),r)} \right\} } \cdot \mathbb{P}\{v\} &&\\ \nonumber
 		&=  \sum_{v'\in R_q} \mathbbm{1}_{ \left\{ \substack{r=\text{HelpRec}(v')\\ \mathbf{k}'=\text{Rec}(v',r)} \right\} } \cdot \mathbb{P}\{v'\} &&\\ \nonumber
 		&= \sum_{v'\in R_q} \mathbb{P}\{v'\} \cdot \mathbb{P}\{ \mathbf{k}' \mid r, v' \} = \mathbb{P}\{ \mathbf{k}' \mid r \}. 
\tag*{\IEEEQED} 
\end{flalign*} 
\let\IEEEQED\relax%
\end{IEEEproof}
Returning to Game 2 and Game 3, we construct an efficient reduction $\mathcal{S}$ as follows: it takes as input two pairs $(\mathbf{A},\mathbf{u}), (\mathbf{b},v)$, and outputs
%\vspace{-1mm}
$$ \left(\mathbf{A}, \mathbf{b}, \;\left(\mathbf{u}, \; r=\text{HelpRec}(v) \right), \; \mathbf{k}=\text{Rec}(v,r)\right).$$
%\vspace{-1.5mm}
%where $\theta$ is the $R$-isomorphism between $R_q$ and $R_q^{\vee}$. 
After that, we will take two indistinguishable inputs, and hence, by efficiency of $\mathcal{S}$, get two indistinguishable outputs.\\
First suppose that the inputs are $M$-LWE instances; i.e. $\mathbf{u}=\mathbf{A}^T\mathbf{s'}+\mathbf{e'}$ and $v=\mathbf{b} \cdot \mathbf{s'}+e''$. So $\mathbf{A}$ must be uniformly random, and $\mathbf{b}$ is indistinguishable from uniform. Hence, the output of $\mathcal{S}$ will be exactly as in Game 2. Now suppose that the inputs given to $\mathcal{S}$ are uniformly random in and independent, then the outputs of $\mathcal{S}$ 
are exactly as in Game 3. In fact, $\mathbf{A},\mathbf{b},\mathbf{u},v$ are uniform, and hence by Theorem \ref{claim peikert}, $\mathbf{k}$ is uniformly random conditioned on $r=\text{HelpRec}(v)$.

To show that Game 3 and Game 1' are %computationally 
indistinguishable, we modify Game 1 and Game 2 by choosing $\mathbf{k}^* \xleftarrow{\text{\$}} \Lambda_2/\Lambda_3$ and output it instead of $\mathbf{k}$.  In this case Game 1 becomes Game 1'. Let Game 2' be the modified version of Game 2. By the same reasoning as above, we can prove that Game 1' is computationally indistinguishable from Game 2' and Game 3.
\begin{rem}
Following the steps in \cite[Section 5]{PEIKERT}, we can construct a passively secure encryption scheme based on
our passively secure KEM, which yields an actively secure encryption scheme and an actively secure key transport protocol.
\end{rem} 

\section{Security against known attacks}
\label{cbd}
We study the hardness of Module-LWE by considering it as an LWE problem, since, to date, the best known attacks don't make use of the module structure.
There are numerous attacks to consider, however, we essentially deal with two BKZ attacks, referred to as primal and dual attacks (see \cite{NEWHOPE, SABER, concrete_hardness} for details). The cost of the primal attack and dual attack are given in Table \ref{security_table} using NewHope's script \footnote{\url{https://github.com/newhopecrypto/newhope/blob/master/scripts/PQsecurity.py}} to do the calculations. We also make a comparison between our protocol and \cite{kyber, kyber3, SABER}'s in the term of security, and obtain a significantly improved security with respect to \cite{kyber, kyber3} with smaller decryption error rate, and the same level of security with better modulus and smaller error rate comparing to \cite{SABER}. 
\begin{table}[h] 
\begin{center}
\begin{tabular}{ |p{1cm}||p{0.6cm}|p{0.6cm}||p{1.4cm} p{1.4cm} p{1.1cm}| }
 \multicolumn{6}{c}{} \\ % This is just to add space above the table
 \hline
 Attack &  $m$ & $b$ & Known Classical & Known Quantum & Best Plausible \\
\hline
\hline
%%%%%%%%%%%%%
% \multicolumn{6}{|c|}{NewHope: $q=12289$, $n=1024$, $k=16$, $P_e \leq 2^{-61}$} \\
 %\hline
%Primal & $1100$ & $967$ & $282$ & $256$ & $200$ \\
% \hline
% & $1099$ & $962$ & $281$ &  $255$ & $199$ \\
% \hline
 %%%%%%%%%%%%
 \multicolumn{6}{|c|}{Saber-KEM: $q=2^{13}=8192$, $n=256$, $k=8$, $d=3$, $P_e \leq 2^{-136}$} \\
  \hline
Primal & $765$ & $667$ & $195$ & $176$ & $138$ \\
 \hline
Dual & $765$ & $664$ & $194$ &  $176$ & $137$ \\
 \hline
 \hline
 %%%%%%%%%%%%%%
\multicolumn{6}{|c|}{Kyber768 Round 1: $q=7681$, $n=256$, $k=4$, $d=3$, $P_e \leq 2^{-142}$} \\
  \hline
Primal & $714$ & $613$ & $179$ & $162$ & $127$ \\
 \hline
Dual & $733$ & $610$ & $178$ &  $161$ & $126$ \\
 \hline
 %%%%%%%%%%%%
\multicolumn{6}{|c|}{Our Protocol: $q=2^{12}=4096$, $n=256$, $k=4$, $d=3$, $P_e \leq 2^{-152}$} \\
  \hline
Primal & $730$ & $667$ & $195$ & $177$ & $138$ \\
 \hline
Dual & $727$ & $664$ & $194$ &  $176$ & $137$ \\
 \hline
 \hline
%%%%%%%%%%%%%%
\multicolumn{6}{|c|}{Kyber768 Round 3: $q=3329$, $n=256$, $k=2$, $d=3$, $P_e \leq 2^{-164}$} \\
  \hline
Primal & $658$ & $623$ & $182$ & $165$ & $129$ \\
 \hline
Dual & $670$ & $620$ & $181$ &  $164$ & $128$ \\
 \hline
 %%%%%%%%%%%%
 \multicolumn{6}{|c|}{Our Protocol: $q=2^{11}=2048$, $n=256$, $k=2$, $d=3$, $P_e \leq 2^{-174}$} \\
  \hline
Primal & $658$ & $665$ & $194$ & $176$ & $138$ \\
 \hline
Dual & $651$ & $662$ & $194$ &  $176$ & $137$ \\
 \hline
\end{tabular}
\end{center}
\caption{ \normalfont  Core hardness of our protocol and comparison with the state of the art. $b$ denotes the block dimension of BKZ, and $m$ the number of used samples. The given costs are the smallest ones for all possible choices of $m$ and $b$. \label{security_table}}
\vspace{1mm}
\end{table}
\section*{Acknowledgments}
The work of C. Saliba and L. Luzzi is supported by the INEX Paris-Seine AAP 2017. The authors would like to thank J.-P. Tillich for helpful comments.
%%%%%%%%%%%%%%%%%%%%%%%%%%%%%%%%%%%%%%%%%%%%%%%%%%%%%%%%
%%%%%%%%%%%%%%%%%%%%%%%%%%%%%%%%%%%%%%%%%%%%%%%%%%%%%%%%%%%%
%\end{appendices}

{\footnotesize

\bibliographystyle{IEEEtran}
\bibliography{References}
}

\end{document}